\documentstyle[aps,prl,preprint]{revtex}
\begin{document}
\draft
\preprint{Nordita 95/81 S, Applied Physics Reports 95-44}
\title{Fluctuating loops and glassy dynamics of a pinned line in two
dimensions}
\author{Anders B.\ Eriksson$^1$, Jari M.\ Kinaret$^2$
\and Lev V.\ Mikheev$^{1}$\cite{bylinelm}}
\address{$^1$ Nordita, Blegdamsvej 17, DK-2100 Copenhagen, Denmark\\
$^2$ Department of Applied Physics, Chalmers University of Technology and 
G\"oteborg University,\\ S-41296 G\"oteborg, Sweden}
\maketitle
\begin{abstract}
We represent
the slow, glassy equilibrium dynamics of a line in a two-dimensional random 
potential landscape as driven by an array of asymptotically
independent two-state systems, or
 {\em loops}, fluctuating on all length scales. 
The assumption of independence enables a fairly complete 
analytic description. We obtain good agreement with Monte Carlo 
simulations 
when the free energy barriers separating the two sides of a 
loop of size $L$ are drawn from a distribution whose width and mean scale 
as  $L^{1/3}$, in agreement with recent results for scaling of such
barriers.
\end{abstract}
\date{\today}

\narrowtext

Slow dynamics is perhaps the most significant characteristic of the 
glassy state of matter, affecting essentially all experimental measurements.
An intuitively appealing picture, which explains this remarkable slowing down,
is that the configuration space of a glass consists of many nearly
degenerate free energy minima, separated by high potential 
barriers \cite{Parisi}.
The dynamics is dominated by transitions between configurations 
whose free energy difference $\Delta E$ 
is less than  the thermal energy $k_BT$, and which are separated 
by free energy barriers $E_B \gg k_BT$. 
In a genuine glass such degeneracies occur 
on all length scales $L$. 
The dependence of $\Delta E$ and $E_B$, or 
more precisely, of their probability distributions, on the linear 
extent $L$ has become the focus of great theoretical interest
\cite{Parisi,FisherHuseSG,HuseHenley,HuHeFi,Ioffe,Fisher2Huse,FisherHuse,MDK}. 
If transitions mainly occur between pairs of low-energy configurations
which can be regarded asymptotically independent for large $L$, the simple 
model of a gas of fluctuating two-level systems \cite{HalperinVarma} 
allows for a fairly complete description of the dynamics. In general, 
however, a much more complicated hierarchical interdependence of 
transitions on different length scales may take place \cite{Parisi}.

The subject of this Letter is an elastic line (henceforth called an
interface) in a 
two-dimensional random potential landscape 
\cite{HuseHenley,FisherHuse,MDK}. This system combines a fair degree 
of realism e.g.\ as a model for a domain boundary \cite{Ioffe} 
or a magnetic flux line trapped between 
two copper-oxide planes in a dirty high-temperature superconductor 
\cite{Fisher2Huse,MDK}, with a simplicity 
that has allowed a substantial body of knowledge to accumulate over the 
past decade \cite{KardarRev,Zhang95}. 
An almost degenerate two-level system in this case is 
simply a {\em loop}: a segment of the line between two points, which 
can flip between two low energy paths (valleys in the potential
landscape) separated by a barrier 
(a mountain in the landscape), this is illustrated in the inset of
Fig.~\ref{fig:tau_stat}. It has been well 
established \cite{HuseHenley,HuHeFi,FisherHuse} 
that the transverse size of such a loop 
scales as $\Delta h\propto L^{\zeta}$ with $\zeta = 2/3$, while the 
free energy differences between the two valley configurations, 
$\Delta E$
are distributed with mean zero and variance 
$\langle\Delta E^2\rangle\propto L^{2\theta}$, 
where $\theta=1/3$. Recent work \cite{MDK} 
provides evidence that the barriers between such configurations 
are distributed with mean $\langle E_B(L)\rangle\propto L^{\theta}$ and the 
variance $\langle E_B^2\rangle \propto L^{2\theta}$ 
with likely logarithmic corrections.

These results, when combined, provide all necessary ingredients for
developing the dynamic description of the line under the assumption
of asymptotic 
independence of the transitions within large loops. In this 
Letter we outline such a description, and show that it agrees 
well with dynamical Monte Carlo simulations. 
We first numerically confirm that nearly degenerate
paths form loops of various sizes, and at first stage we analyze the 
dynamics of loops of a fixed length. The dynamics of a single loop
can be described as a sequence of flips where the interface position
moves from one arm of the loop to another. By studying
loops of different lengths we determine how the flipping-rate distribution
depends on the size of the loop. At the second stage we study the fluctuations
of a low-energy interface. Using numerical simulations we determine the 
time-dependent fluctuations of the interface position, which we compare
with an analytic model that assumes that interface fluctuations are due to
flipping loops. Since parameters describing loops are determined 
at the first stage, there are no adjustable parameters left at this point. 
We find that
the prediction of the loop model is in good agreement with the simulations,
supporting the conjecture that interface dynamics is due to fluctuating loops.

\paragraph*{The numerical model:}
We study a model that excludes overhangs so that the interface height
$h(x)$ is at all times a single-valued function of the spatial coordinate
$x$. For the numerics we use a lattice model where the interface is
discretized to have a unit slope between the lattice points, 
$|h(x)-h(x+1)| = 1$, and use fixed boundary conditions $h(0)=h(L_0)=0$,
where $L_0$ is the interface length.
The Hamiltonian of the lattice model for a particular
realization of the random medium is 
\begin{equation}
H_\mu[h] = \sum_{x=1}^{L_0} \mu(h(x),x) ,
\end{equation}
where $h(x), x=1,\ldots,L_0$ represents the interface, and
$\mu$ is the potential landscape. The random
potential is uncorrelated and uniformly distributed over the range
$0 \leq \mu(h,x) \leq 1$ \cite{Gaussian}.
This  Hamiltonian belongs to the universality class of a directed polymer in a
random medium (DPRM) \cite{Zhang95,Yoshino95}, and in this context the
requirement of unit
slope between lattice points effectively adds an implicit line tension
\cite{Zhang95}. The dynamics is implemented in a spirit similar to
that of previously introduced mappings onto spin chains
\cite{GwaSpohn,HansAndersLev}; the
dynamics is described by the master equation based on the transition
probability $P[h(x)\to h'(x)] = \exp\{\beta(H_{\mu}[h]-H_{\mu}[h'])/2\} dt$,
 where $\beta$ is the inverse temperature, $dt$ an infinitesimal time
interval, and $h(x)$ and $h'(x)$ are two interface configurations that
differ in only one position. Numerically the dynamics of
the master equation is exactly modeled by an algorithm that uses time
steps sampled from a Poisson distribution \cite{Binder79}.
We have used $\beta=2$ in the simulations that are presented in this Letter.
In general
the computation time grows very rapidly with increased $\beta$.

\paragraph*{Single loop statistics:}
In order to explore the ideas of loop-based dynamics we use the free
energy landscape to define loops.
The free energy of a point $(x,h)$ is
defined as $F(x,h) = -\beta^{-1} \ln(P(x,h))$, where $P(x,h)$ is the
probability that an interface with fixed ends ($(x,h)=$ $(0,0)$ and
$(L_0,0)$) crosses this point. The probability $P(x,h)$ is
easily calculated using transfer matrix methods \cite{Zhang95}. A
lattice point is defined
to be part of an {\em island} if it is not part of any interface that
includes
only points with $P(x,h)>0.1$. The level $0.1$ is chosen to
obtain well defined islands, but the specific value does not greatly affect
the final results of the analysis. Interface segments
encircling an island form a loop, and we 
measure the loop size in terms
of its length $L=x_r-x_l+1$, where $(x_r,h_r)$ and $(x_l,h_l)$
are the right and left ends of the island, respectively. 
The center  height of the
island, $h_{is}$, is defined as the $h$-component of its center of
mass, where  every
lattice point in the island is regarded as a mass point. 
We say that the loop surrounding the island changes its state (flips) when
the interface height, averaged over the island length, crosses $h_{is}$.
This definition is convenient although it sometimes  catches  events
that would not intuitively be considered  as flips.

In order to study the relationship between static and dynamic scaling,
and to unambiguously determine the numerical values of the amplitudes to
be later used in the fit, we
collect statistics of the dynamics of individual loops. A loop is
inscribed in a bounding box which sets the boundaries within which the
interface is free to move (showed with dashed lines in
Fig.~\ref{fig:tau_stat}). The interface is
constrained to pass
through the left and right corners of the bounding box,
$(x_{l}-2,h_{l})$,
and $(x_{r}+2,h_{r})$. We collect statistics of the time between
consecutive flips of the loop.
In Fig.~\ref{fig:tau_stat} we plot a typical example of the measured 
probability that a loop stays in the upper (lower) state for at least 
time $t$.
The decay of this probability is exponential 
for sufficiently large $t$,
which is consistent with the two-state model discussed below 
(Eq.~(\ref{eq:twostate})). The deviation
for small $t$ is probably due to our simplified criterion for flipping
the loop. Before
making a least-square fit to an exponential decay we exclude the
part of the short-time data that strongly deviates from the expected form. 
We also exclude the largest times (0.1\% of the data)
since the statistics of these very rare events is poor. From this we
obtain the characteristic
decay times $\tau_{+-}$ and $\tau_{-+}$ for flipping the loop from
its upper to lower state and vice versa. 

For an individual loop we collect data from 10,000
flips, and make the fit described above. 
The rate constant,
$\Gamma$, characteristic for the loop, is calculated by
$\Gamma=\tau_{+-}^{-1}+\tau_{-+}^{-1}$
(we assume that the free energy difference between the arms of the
loop is unimportant). By collecting statistics from 1000
loops of the same size $L$ we find that $\Gamma$
is log-normally distributed. This is consistent with a
simple activated behavior, $\Gamma =
\overline{\Gamma}e^{-\beta\Delta}$, where $\Delta$ is a
Gaussian-distributed energy barrier separating the two sides of the
loop, and $\overline{\Gamma}$ is a constant setting the unit of
time. In Fig.~\ref{fig:ln} we show fits to
the log-normal distribution by plotting $Q^{-1}(P(\Gamma))$
against $\ln(\Gamma)$, where $Q$ is the complement of the cumulative
normal distribution,
and $P(\Gamma)$ is the measured cumulative probability of the rate
constant $\Gamma$ (hence, a straight line would correspond to a
log-normal distribution). 
The distribution is characterized by the average barrier height $\Delta(L)$
and the standard deviation $\sigma(L)$.
The values for $\beta\sigma(L)$ are obtained as the standard deviations of
$\ln(\Gamma)$, which also give the inverse slopes of the lines
in Fig.~\ref{fig:ln}. Similarly, the crossings between the lines and
the zero axis represent the averages of $\ln(\Gamma)$, which give values
for  $\ln(\overline{\Gamma}) - \beta\Delta(L)$. The
average barrier height $\Delta(L)$ and the standard deviation $\sigma(L)$
scale with loop size as $L^{1/3}$ \cite{MDK}.
This allows us to determine the time scale 
$\overline{\Gamma}$ and the multiplicative
constants in $\Delta(L)$ and $\sigma(L)$. The mean and standard deviation
of $\ln(\Gamma)$ are plotted as functions of $L^{1/3}$ 
in Fig.~\ref{fig:exp_1_3} confirming the $L^{1/3}$ scaling,
and giving for $\beta=2$ the parameter values
$\ln(\overline{\Gamma}) = 2.2$,
$\beta\Delta(L) =
3.0 L^{1/3}$, and $\beta\sigma(L) = 0.28 L^{1/3}$.

\paragraph*{Full interface numerics:}
The same numerical algorithm is used for studying the dynamics of the
full interface but keeping only the ends of 
the interface fixed. We start the interface from a random initial state
(chosen from the equilibrium ensemble),
measure how the height $h_c(t)$ of the center point of the
interface varies with time, and collect statistics of $\delta h^2(t) =
[h_c(t)-h_c(0)]^2$.
The equilibrium ensemble is used to normalize $\delta h^2(t)$
with respect to $\delta h^2(\infty)$. The data points in 
Fig.~\ref{fig:dh2_of_t} show the results of simulating interfaces
of lengths 20,40,60, and 128 for $\beta=2$, where $\delta h^2(t)$ has
been averaged over 20,000 realizations of the random medium.

\paragraph*{The analytic model:}
We model the loop dynamics using a two-state model, where the probability
of a given loop being in the upper or lower state, respectively, is given by 
$P_+(t)$ and $P_-(t)$. The time development of these probabilities is governed
by the coupled differential equations
\begin{equation}
\begin{array}{lcl}
\frac{dP_+}{dt}
 &=&
 -\Gamma_{+-}P_{+}(t) + \Gamma_{-+}P_{-}(t) ,
\vspace{1mm}\\
\frac{dP_-}{dt}
 &=&
 \Gamma_{+-}P_{+}(t) - \Gamma_{-+}P_{-}(t) .
\end{array} 
\label{eq:twostate}
\end{equation}
The fluctuation
in the height of the center position of a loop of width $wL^{2/3}$ 
is given in the two-state model by 
\begin{equation}
\delta h_{\Gamma,L}^2(t) 
= \langle \left[h_c(t) - h_c(0)\right]^2\rangle
= \frac{w^2}{2}L^{4/3}(1 - e^{-\Gamma t}) ,
\end{equation}
where we assumed for simplicity 
$\Gamma_{+-} = \Gamma_{-+} = \Gamma/2$, i.e.\ that the
two arms of the loop are exactly degenerate. 

We consider an interface of length $L_0$, 
and denote the average barrier height of the
largest loops (i.e.\ loops of size $L_0$) 
by $\Delta_0$, and its standard deviation by $\sigma_0$.
We assume that the number of loops of a given size scales as $L^{-1}$,
and that the fluctuations of different loops are independent and additive.
We find that the total fluctuation of the interface, as implied by
loop dynamics, is given by
\begin{equation}
\label{eq:result}
\frac{\delta h^2(t)}{\delta h^2(\infty)} = 1 - \int_{-\infty}^\infty
\frac{dy}{\sqrt{2\pi}}e^{-\frac{1}{2}y^2}
\int_0^1du4u^3
e^{-\overline{\Gamma}t\exp\left[u\beta(\sigma_0y-\Delta_0)\right]} .
\end{equation}
The first integral corresponds to integrating over
loops of fixed length
but variable rate constants, and the second integral adds up the 
contributions of loops of different lengths. The infinite time fluctuations
$\delta h^2(\infty)$ are given by the equilibrium result, and scale
as $L_0^{4/3}$ \cite{FisherHuse}.

The two-state model (Eq.~(\ref{eq:result}))
implies that the natural time scale in the 
problem is given by
$\overline{t}=(\overline{\Gamma})^{-1}\ln(2)\exp(\beta\Delta_0/2^{1/4})$
in the sense that at time $t=\overline{t}$ the fluctuations have reached
approximately 50\% of the equilibrium value; more precisely
$0.5 \le \frac{\delta h^2(\overline{t})}{\delta h^2(\infty)} \le 0.58$
for all $\beta$, $\Delta_0$, and $\sigma_0$. Hence, at low temperatures
interface dynamics are exponentially slow as is typical for glassy systems
\cite{Hertz}.

The solid lines in Fig.~\ref{fig:dh2_of_t} 
are the results of the two-state model using the parameters determined
by studying individual loops. 
The time scale $\overline{\Gamma}$ determines the position of the
curves, the average barrier height, $\Delta_0$, 
affects both the position and the
slope of the curves, and the standard deviation of the barrier heights,
$\sigma_0$, has only a minor effect on the results for small
$\sigma_0/\Delta_0$
(taking the limit $\sigma_0/\Delta_0\rightarrow 0$ would
not significantly change the results).
 Considering that there are no adjustable
parameters, the agreement with the results of the numerical simulations
in Fig.~\ref{fig:dh2_of_t} is quite good. 
This suggests that the hypothesis 
that interface dynamics is due to fluctuating loops is indeed valid. 
The deviations in the short time behavior are, we believe, due in part to 
more complicated loops on small length scales, which cannot be
described as independent simple loops. Another contribution to the short
time deviations is that in the lattice model the arms of a loop can have
a non-zero width (i.e.\ two adjacent interface positions are not separated
by a barrier), which naturally enhances fluctuations in short time scales.
The interpretation that small time differences are due to discretization
and finite size corrections is further supported by the fact that the 
deviations are smaller for larger system sizes.
Another possible source of deviations is logarithmic corrections to the
scaling forms of $\Delta(L)$ and $\sigma(L)$,
however, our data on individual loops is insufficient to determine
these corrections.

In conclusion, we have studied the dynamics of a one-dimensional interface
in a two-dimensional random medium. We have shown that the dynamics can be
understood quantitatively in terms of loops formed by nearly degenerate
interface paths. At low temperatures the dynamics is exponentially slow,
which is typical for glassy systems \cite{Parisi,FisherHuseSG,Hertz}.

\input{epsf}
\newlength{\figurewidth}
\setlength{\figurewidth}{10.5cm}

\begin{figure}
\centerline{
\epsfxsize=\figurewidth
\epsfbox{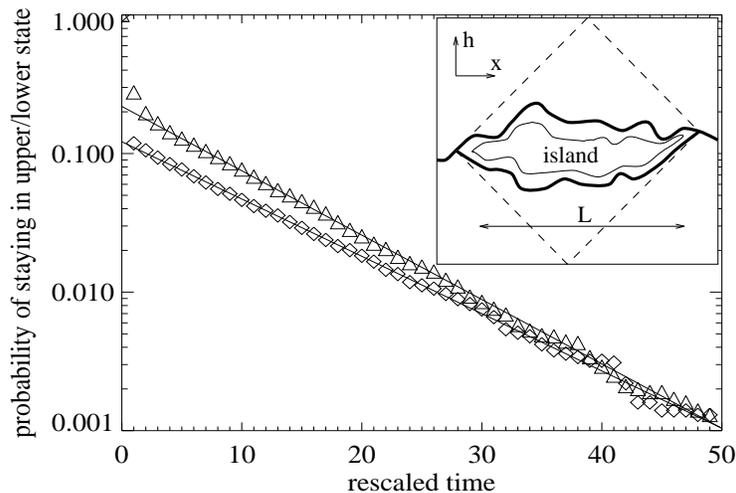}
}
\caption{Typical time dependence  of the probability that a loop stays in the
upper (lower) state for at least a time $t$.
The data is based on 10,000 crossings in
each direction for an island of size $L=10$. Note that the time
scales are normalized to make the right-end value
0.1\% of the initial value. The inset schematically shows a  loop formed
by two nearly degenerate alternative paths of the interface.  }
\label{fig:tau_stat}
\end{figure}

\begin{figure}
\centerline{
\epsfxsize=\figurewidth
\epsfbox{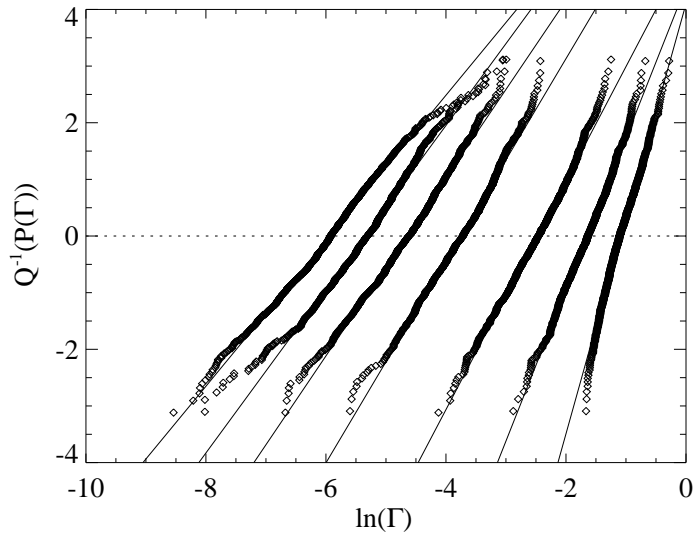}
}
\caption{The cumulative distribution, $P(\ln(\Gamma))$, and the cumulative
normal distribution $Q$ are used to illustrate the fit to a log-normal
distribution of the $\Gamma$ data (solid lines) for the loop sizes
$L=1,2,4,8,12,16,20$ (from right to left). The statistical data is based
on 1000 loops of each size.
}
\label{fig:ln}
\end{figure}

\begin{figure}
\centerline{
\epsfxsize=\figurewidth
\epsfbox{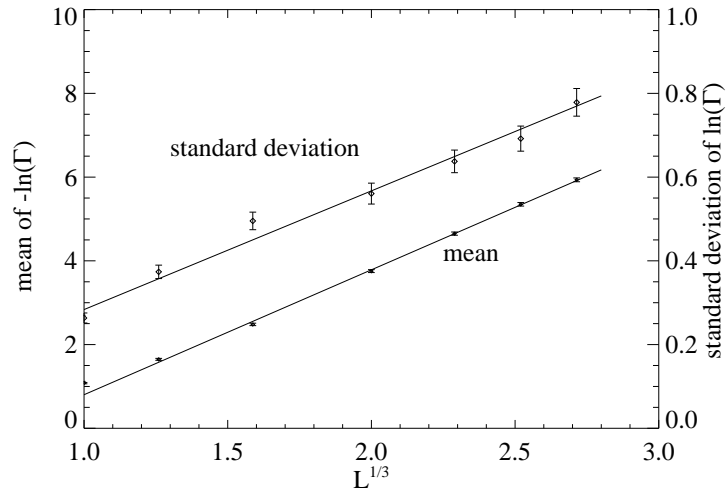}
}
\caption{The parameters $\Delta(L)$, $\sigma(L)$, and $\overline{\Gamma}$
are determined by fitting $\beta\Delta(L)-\ln(\overline{\Gamma})$ to the
mean, and $\beta\sigma(L)$ to the standard deviation of the
$\ln(\Gamma)$ data. The error bars show the 95\% confidence
interval of the statistical error.}
\label{fig:exp_1_3}
\end{figure}

\begin{figure}
\centerline{
\epsfxsize=\figurewidth
\epsfbox{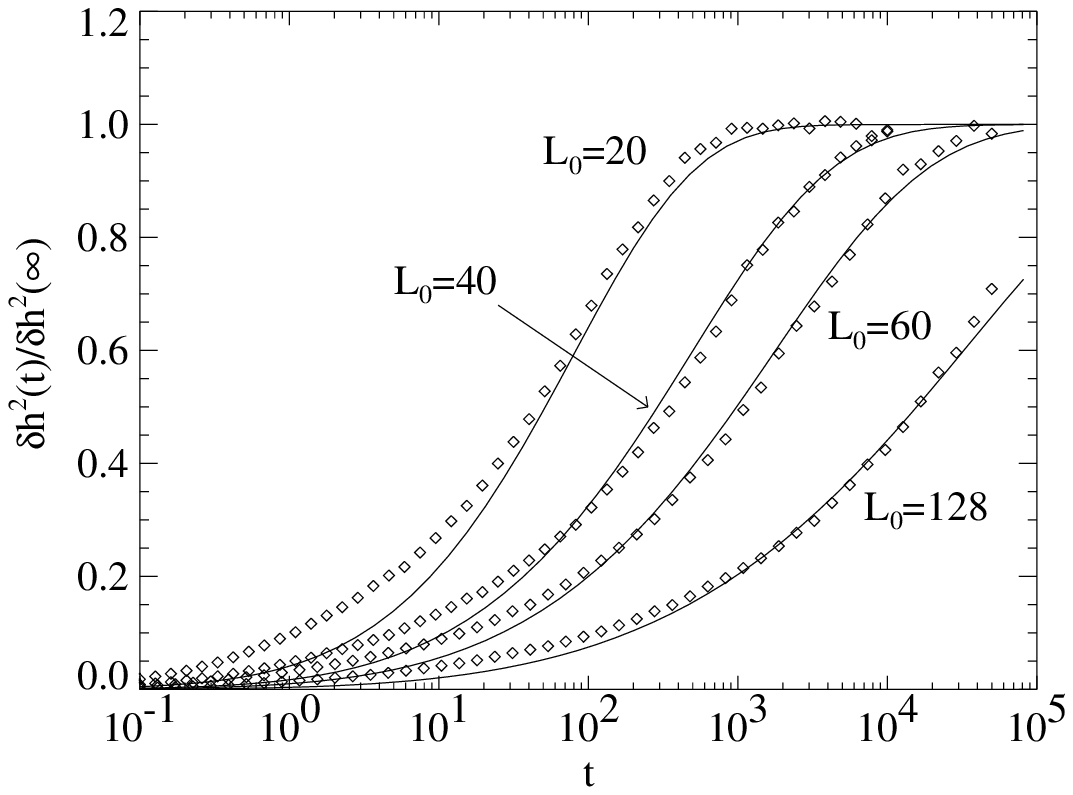}
}
\caption{The points show the time dependence of $\delta h^2(t)$ for
interfaces of lengths 20, 40, 60, and 128 averaged over 20,000 samples each.
The solid lines show the result of the loop model
(Eq.~(\protect\ref{eq:result})) where the
parameters of the single-island analysis have been used.
}
\label{fig:dh2_of_t}
\end{figure}


\begin{references}{}
\bibitem{Gaussian}
A Gaussian distribution with the same mean and standard
deviation gives results that marginally differ from those of the
uniform distribution.
\end{references}
\end{document}